\newif\ifsingle

\documentclass[a4paper,conference]{IEEEtran}
\IEEEoverridecommandlockouts

\usepackage[top=1.9cm, bottom=4.45cm,left=1.3cm,right=1.34cm]{geometry}
\def\BibTeX{{\rm B\kern-.05em{\sc i\kern-.025em b}\kern-.08em T\kern-.1667em\lower.7ex\hbox{E}\kern-.125emX}}
\usepackage[noadjust]{cite}
\usepackage[T1]{fontenc}
\usepackage[latin9]{inputenc}
\usepackage{color}
\usepackage{amsmath}
\usepackage{amssymb}
\usepackage{graphicx}
\usepackage{textcomp}
\usepackage{xcolor}
\usepackage[bookmarks,hidelinks]{hyperref}

\makeatletter

\providecommand{\tabularnewline}{\\}

\ifsingle
\newcommand{\figWidth}{0.65\columnwidth} 
\else
\newcommand{\figWidth}{1\columnwidth} 
\fi 

\definecolor{NewColor}{rgb}{0.2,0,0.5}

\makeatother

\begin{document}
\title{Deep Learning-Based Signal Detection for Dual-Mode Index Modulation 3D-OFDM}

\author{Dang-Y Hoang, Tien-Hoa Nguyen, Vu-Duc Ngo, Trung Tan Nguyen, Nguyen Cong Luong, and Thien Van Luong

\thanks{Dang-Y Hoang, Tien-Hoa Nguyen, and Vu-Duc Ngo are with the School of Electrical and Electronics Engineering, Hanoi University of Science and Technology, Hanoi, Vietnam (e-mail: hoangy211099@gmail.com, \{hoa.nguyentien,duc.ngovu \}@hust.edu.vn.)}

\thanks{Trung Tan Nguyen is with the Faculty of Radio-Electronics, Le Quy Don Technical University, Ha Noi 11355, Vietnam (e-mail: trungtannguyen@mta.edu.vn).}

\thanks{Thien Van Luong and Nguyen Cong Luong are with the Faculty of Computer Science, Phenikaa University, Hanoi 12116, Vietnam (e-mail: \{thien.luongvan, luong.nguyencong\}@phenikaa-uni.edu.vn). }




\vspace{-0.5cm}
 }

\maketitle
\vspace{-1cm}

\begin{abstract}
In this paper, we propose a deep learning-based signal detector called DuaIM-3DNet for dual-mode index modulation-based three-dimensional (3D) orthogonal frequency division multiplexing (DM-IM-3D-OFDM). Herein, DM-IM-3D-OFDM is a subcarrier index modulation scheme which conveys data bits via both dual-mode 3D constellation symbols and indices of active subcarriers. Thus, this scheme obtains better error performance than the existing IM schemes when using the conventional maximum likelihood (ML) detector, which, however, suffers from high computational complexity, especially when the system parameters increase. In order to address this fundamental issue, we propose the usage of a deep neural network (DNN) at the receiver to jointly and reliably detect both symbols and index bits of DM-IM-3D-OFDM under Rayleigh fading channels in a data-driven manner. Simulation results demonstrate that our proposed DNN detector achieves near-optimal performance at significantly lower runtime complexity compared to the ML detector.
\end{abstract}


\vspace{-0.2cm}

\section{Introduction}
Orthogonal frequency division multiplexing (OFDM) with index modulation OFDM-IM \cite{barsar2013ofdmim,ThienTVT2017} has recently materialized as a promising multicarrier modulation candidate to substitute the traditional OFDM. In contrast to OFDM, only a portion of subcarriers are activated in OFDM-IM, where additional information bits are virtually carried using the indices of activated sub-carriers, while other information bits are physically conveyed on the classical $M$-ary symbols modulated over these activated sub-carriers. As a result, OFDM-IM outperforms the conventional OFDM in terms of reliability and energy efficiency, particularly a promising trade-off between reliability and spectral efficiency performance through adjusting the number of activated sub-carriers.

A variety of cutting-edge IM-based systems have been developed over the years. For instance, a tight bound on the bit error rate (BER) of the low-complexity greedy detector was derived in \cite{thien2017MCIK}, while study in \cite{ThienTVT2017} examined its symbol error probability (SEP) when using different detection types such as the maximum likelihood (ML) and GD detectors. The outage probability of OFDM-IM using the GD detector was first analyzed in \cite{Pout2017}. The combination of OFDM-IM with multiple antennas was introduced in \cite{Basar2016mimoIM}. The BER performance of OFDM-IM in presence of maximal-ratio combiner (MRC) for enhancing diversity was investigated in \cite{thienWCL2018}. Since then, a range of techniques has been applied to either increase the spectral efficiency (SE) or reduce the BER of the OFDM-IM system. In particular, in order to improve SE, the IM concept is applied to both in-phase and quadrature (I/Q) parts for doubling the number of index bits. A low-complexity ML detector and the performance analysis for this system were also reported. Additionally, the dual mode OFDM (DM-OFDM), which employs several constellations to transmit data bits using inactive subcarriers of OFDM-IM, was proposed in \cite{Mao2017dmIM}. It was demonstrated that at the penalty of degraded error performance, DM-OFDM can attain higher SE than OFDM-IM. In \cite{wen2017mmIM}, a generalized variant of DM-OFDM that vastly improves the SE over the original version was proposed. Other methods for increasing diversity of OFDM-IM were reported in \cite{ThienTWC2018,wen2017enhancedIM,ThienTVT2018}, which considerably improve BER performance, especially in high signal-to-noise ratio (SNR) regions. 

Aiming at improving the error performance of the classical OFDM, three-dimensional (3D) constellation and OFDM were combined to form 3D-OFDM in \cite{kang20083dofdm}. With 3D-OFDM, data bits are mapped into a 3D constellation and the 3D signals are divided into sub-carriers, which are modulated using the 2D inverse fast Fourier transform (IFFT). In comparison with OFDM with the 2D mapper, the 3D mapper considerably increases the minimum Euclidean distance (MED) between the various constellations, leading to improved error performance. In \cite{chen2010clo} and \cite{huang2018papr}, the theoretical SEP analysis and the peak-to-average power ratio (PAPR) reduction method of the 3D-OFDM were respectively presented. However, when the constellation order, SE of 3D-OFDM is only one-third that of traditional OFDM \cite{CHEN2020impro}. Inspired by dual-mode OFDM \cite{Mao2017dmIM}, a novel IM-based scheme called dual-mode index modulation assisted 3D-OFDM (DM-IM-3D-OFDM) was proposed in \cite{Wang2021dm3d} to improve SE of 3D-OFDM. In this scheme, subcarriers are separated into several sub-blocks and subcarriers in each sub-block are partitioned into two groups, then modulated by designed dual-mode 3-D constellations based on the floor of the Poincare sphere \cite{Chen2011}. Compared to 3D-OFDM, DM-IM-3D-OFDM provides better BER performance at high SNRs under fading channels when the optimal ML detector is employed. However, this conventional detector suffers from high computational complexity, especially when the system parameters increase. This paper appears to address this fundamental issue.


In particular, we propose a deep learning-based detector called DuaIM-3DNet for DM-IM-3D-OFDM. More particularly, we design a novel deep neural network (DNN) structure for DuaIM-3DNet which consists of two sub-networks called IndexNet and SymbolNet for detecting active sub-carrier indices and 3D constellation symbols. The outputs of these two sub-networks are then concatenated to be fed into a final DNN transmitted bits using Sigmoid function as the activation of the output layer. Note that before entering the DNN architecture of our proposed DuaIM-3DNet, the received signal and channel state information (CSI) are pre-processed based on the domain knowledge such as zero-forcing (ZF) channel equalizer and energy detection \cite{Thien2019DLIM}. DuaIM-3DNet is trained offline using simulated dataset to minimize BER. Then, the trained model can be deployed as an online detector for promptly detecting transmitted data bits. Finally, our simulation results demonstrate that our DuaIM-3DNet detector can provide a near-optimal error performance at much lower runtime complexity in comparison with the ML detector.

Note that deep learning \cite{SCHMIDHUBER201585} in conjunction with DNNs has been widely adopted to wireless communication systems, particularly in physical layer issues. For example, deep learning was applied for detecting signals of OFDM-IM in \cite{Thien2019DLIM}. Then, a range of studies, which apply deep learning to either improve the error performance or reduce the complexity of OFDM-based multicarrier schemes \cite{Luong2020engery, Chao2022tubro,Luong2020MC-AE,Luong2022optical}. Some other applications of deep learning to channel coding as well as successive interference cancellation receivers of non-orthogonal downlink can be found in \cite{Luping2022Unity} and \cite{Luong2022SIC}, respectively.

The remaining parts of this paper are structured as follows. The DM-IM-3D-OFDM system model is discussed in Section~\ref{sec:System-Model}, while Section~\ref{sec:DeepNet} introduces the proposed DuaIM-3DNet detector. Section~\ref{sec:SIMULATION RESULTS} provides simulation results and discussion. Finally, Section~\ref{sec:Conclusions} concludes our paper.


\vspace{-0.2cm}

\section{DM-IM-3D-OFDM system model\label{sec:System-Model}}

\vspace{-0.1cm}
First, we briefly describe a DM-IM-3D-OFDM system~\cite{Wang2021dm3d}. In DM-IM-3D-OFDM, the information bits are mapped into two different types of 3-D constellations and sub-carrier indices. Assume that $m$ information bits are divided into $u$ sub-blocks. Each sub-block contains $n$ sub-carriers and $p$ bits, i.e. $u=m/p=N/n$, where $N$ is the total number of subcarriers. Then, $p$ bits will be divided into two components. The first one consisting of $p_1$ bits is to map the sub-carrier indices. The second component with $p_2$ bits is mapped by two different constellations of type A and type B, denoted by $\mathcal{C}_A$ and $\mathcal{C}_B$, respectively, $\mathcal{C}_A \cap \mathcal{C}_B = \varnothing$. The number of signal points of $\mathcal{C}_A$ and $\mathcal{C}_B$ are presented by $c_A$ and $c_B$, respectively. We also assume that $k$ stands for the number of sub-carriers mapped to $\mathcal{C}_A$, so the number of sub-carriers mapped to $\mathcal{C}_B$ will be $n-k$. Then, $p_1$ and $p_2$ are calculated as follows
\begin{equation}
    p_1 = \lfloor \log_2C_n^k \rfloor,
\end{equation}
\begin{equation}
    p_2 = \log_2((c_A)^k) + \log_2((c_B)^{n-k}),
\end{equation}
where $C_n^k$ is the binomial coefficient and $\lfloor.\rfloor$ denotes the integer floor operator. The number of information-carrying bits in each sub-block is $p = p_1 + p_2$. An example of the index mapping process is illustrated in Table~\ref{tab:lookup}.
\begin{table}[!ht]
\centering
\caption{Look-up table for index modulation with $n=4, p_1=2 $ and $k=2$\label{tab:lookup}}
\begin{tabular}{|c|c|c|}
\hline
Index Bits & Indices for $\mathcal{C}_A$ & Sub-block \\ \hline
$[0,0]$           & $[1,2]$    & $[A^{(1)},A^{(2)},B^{(1)},B^{(2)}]$                                              \\ \hline
$[0,1]$           & $[2,3]$    & $[B^{(1)},A^{(1)},A^{(2)},B^{(2)}]$                                               \\ \hline
$[1,0]$           & $[3,4]$    & $[B^{(1)},B^{(2)},A^{(1)},A^{(2)}]$                                               \\ \hline
$[1,1]$           & $[1,4]$    & $[A^{(1)},B^{(1)},B^{(2)},A^{(2)}]$                                               \\ \hline
\end{tabular}
\end{table}

\begin{figure}[!ht]
    \centering
    \includegraphics[width=\figWidth]{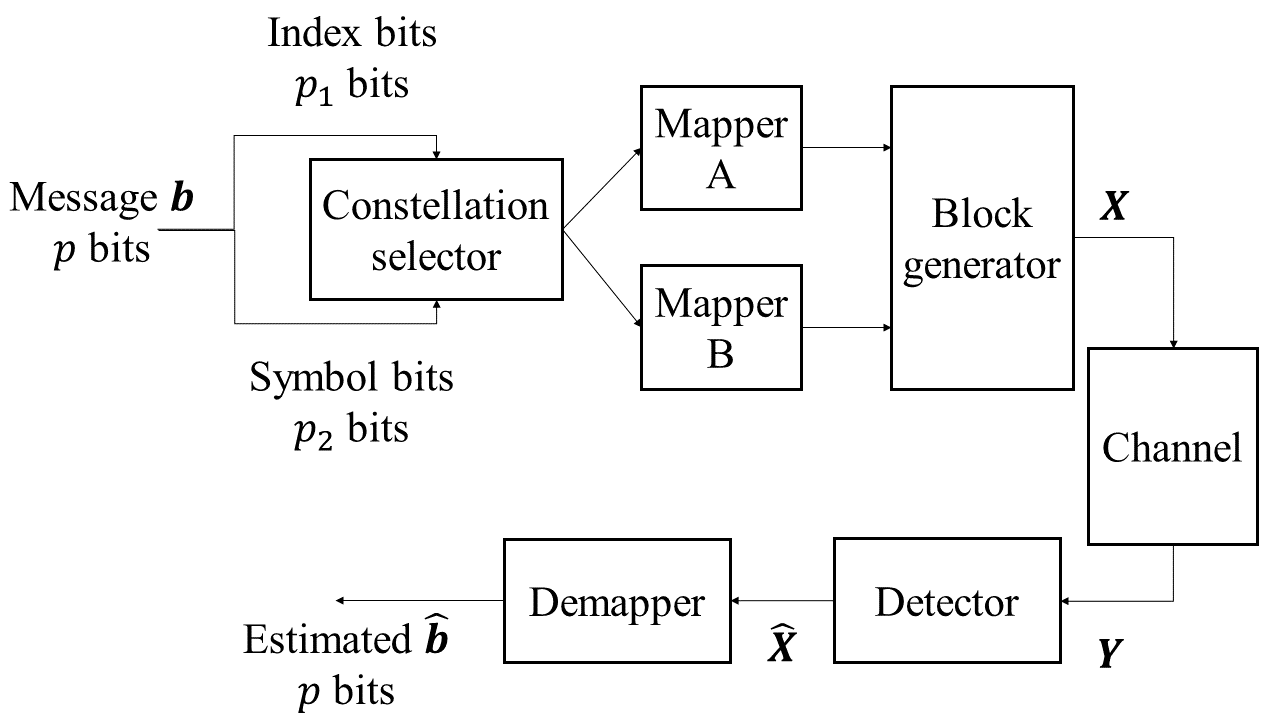}
    \caption[]{Diagram of simple DM-IM-3D-OFDM.}
    \label{fig1}
\end{figure}

Fig.~\ref{fig1} depicts the design for a basic DM-IM-3D-OFDM system with a block of sub-carriers. The symbols transmitted in a block are expressed as the matrix $\mathbf{X} \in \mathbb{C}^{3 \times n} $, which is represented in the frequency domain as $\mathbf{X}=[\mathbf{X}(1),\mathbf{X}(2),...\mathbf{X}(n)]$, where $\mathbf{X}(\varphi) \in \left\{\mathcal{C}_A, \mathcal{C}_B\right\}$, $1 \leq \varphi \leq n$. Note that the design of 3-D constellations for constructing the symbol matrix $\mathbf{X}$ is detailed in \cite{Wang2021dm3d}. The look-up table in Table~\ref{tab:lookup} may be used to create an $\mathbf{X}$ block. For each sub-block, the channel impulse response (CIR) matrix of the Rayleigh fading channel is shown in the frequency domain as follows
\begin{align}    \mathbf{H} = \begin{pmatrix}
          H_{11} & H_{12} & \ldots & H_{1n} \\
          H_{21} & H_{22} & \ldots & H_{2n} \\
          H_{31} & H_{32} & \ldots & H_{3n}
         \end{pmatrix}
         = \begin{pmatrix}
          \mathbf{H}_1\\
          \mathbf{H}_2 \\
          \mathbf{H}_3
         \end{pmatrix},
    \label{congthuc211}
\end{align}
where $\mathbf{H} = [\mathbf{H}_1;\mathbf{H}_2;\mathbf{H}_3]^T \in \mathbb{C}^{3\times n}$ is matrix of size $3\times n$. and the components $\mathbf{H}_i \in \mathbb{C}^{1 \times n}$, $1 \leq i \leq 3$, have the identical and independent distribution with zero mean and unit variance. 

The received signal matrix $\mathbf{Y} \in \mathbb{C}^{3 \times n}$ is performed as follows
\begin{equation}
    \mathbf{Y} = \mathbf{H}\mathbf{X} + \mathbf{Z},
\end{equation}
where $\mathbf{Z}$ denotes the additive white Gaussian noise (AWGN), $\mathbf{Z} \in \mathbb{C}^{3 \times n}$,  $\mathbf{Z}=[\mathbf{Z}_1,\mathbf{Z}_2,\mathbf{Z}_3]^T$ and the components $\mathbf{Z}_j \in \mathbb{C}^{1 \times n}$, $1 \leq j \leq 3$, have the identical and independent distribution with zero-mean and variance of $N_0$.

\vspace{-0.2cm}

\section{Proposed DuaIM-3DNet\label{sec:DeepNet}}

\vspace{-0.1cm}
This section describes the training process and online deployment of the proposed DuaIM-3DNet detector and the neural network structure for our detector is also presented.

\vspace{-0.2cm}
\subsection{Structure of DuaIM-3DNet\label{subsec:Structure of DeepNet}}

\begin{figure}[!ht]
    \centering
    \includegraphics[width=\figWidth]{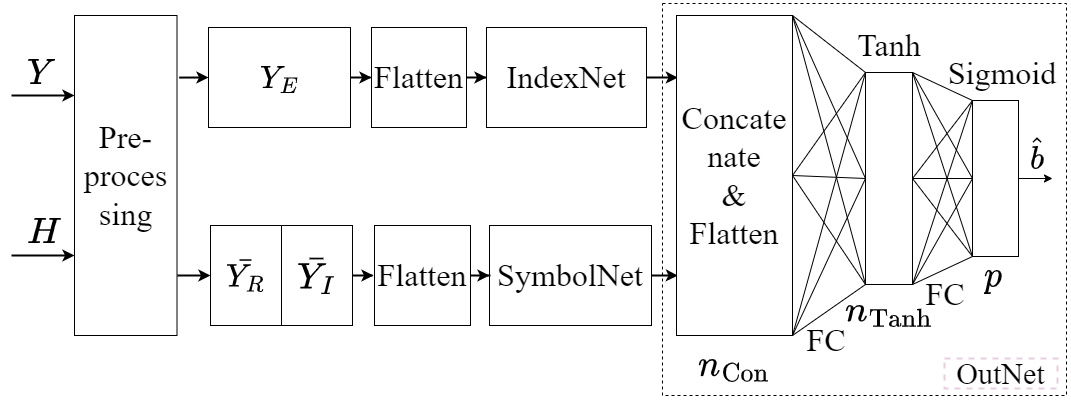}
    \caption[]{Structure of the proposed DuaIM-3DNet.}
    \label{fig2}
\end{figure}

\vspace{-0.1cm}
The proposed DuaIM-3DNet structure is illustrated in Fig.~\ref{fig2} which consists of two identical parallel sub-nets, namely IndexNet and SymbolNet. First, we use a zero-forcing (ZF) channel equalizer to obtain the equalized received signal as $\Bar{\mathbf{Y}}=\mathbf{Y}\mathbf{H}^{-1}$. The channel matrix $\mathbf{H}$ and the received signal matrix $\mathbf{Y}$ are known at the receiver. The IndexNet is responsible for extracting index state characteristics of carriers whose input is a vector, by flattening the received signal's energy matrix represented as $\mathbf{Y}_E$. The SymbolNet is used for demapping the symbol modulation, whose input is the real imaginary part vector by flattening the equalized signal matrix $\Bar{\mathbf{Y}}$, $\Bar{\mathbf{Y}}_R$ and $\Bar{\mathbf{Y}}_I$, respectively. In the second step, outputs of the two sub-nets are concatenated and flattened into a new one-dimensional feature, and then the two FC layers are used to recover the transmitted message $\mathbf{b}$.

\begin{figure}[!ht]
    \centering
    \begin{tabular}{cc}
         \includegraphics[width=0.2\textwidth]{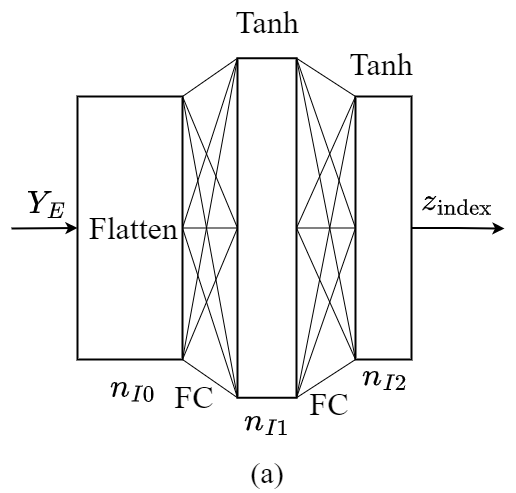} &
         \includegraphics[width=0.2\textwidth]{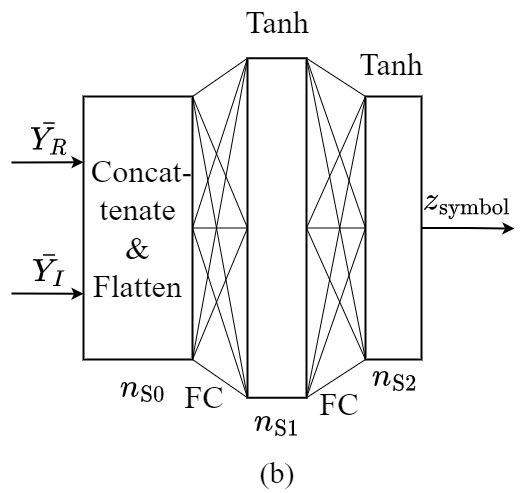}
    \end{tabular}
    \caption[]{Structure of IndexNet (a) and SymbolNet (b).}
    \label{fig3}
\end{figure}
Each sub-net, i.e., IndexNet and SymbolNet, in DuaIM-3DNet consists of two hidden layers. The number of nodes in the output layer is equal to the number of predicted bits in each sub-block, i.e., $n_{\text{out}} = p$. Details of IndexNet and SymbolNet parameters are shown in the Table~\ref{tab:paranet}. In the hidden layer, the Tanh function is used as the activation function $f_{\text{Tanh}}(x)=\frac{e^x-e^{-x}}{e^x+e^{-x}}$. The sigmoid function $f_{\text{Sig}}(x)=\frac{1}{1+e^{-x}}$ is used to map outputs in the interval $(0,1)$. The IndexNet contains two FC layers that use the Tanh function, as illustrated in Fig.~\ref{fig3}(a). The numbers of nodes of two hidden layers are  $n_{\text{I1}}$ and $n_{\text{I2}}$, where $n_{\text{I2}}$ is the number of output nodes of the network and the output of the network is performed by $\mathbf{z}_{\text{index}}$. Similarly, the SymbolNet contains two FC layers whose activation function is Tanh, as depicted
in Fig.~\ref{fig3}(b). It can be seen from the figure that $n_{\text{S1}}$ and $n_{\text{S2}}$ are the numbers of nodes of the FC layers. In particular, $n_{\text{S2}}$ is the number of output nodes of the network, given that the output of the network is performed as $\mathbf{z}_{\text{sym}}$. The output of IndexNet and SymbolNet will be used as input of OutNet block.

\begin{table}
\caption{\textcolor{black}{Parameters of IndexNet, SymbolNet and OutNet with $\left(n,k,c_A,c_B\right)=\left(4,2,2,2\right)$}\label{tab:paranet}}
\centering{}
\begin{tabular}{|l|c|c|c|}
\hline 
Parameter & IndexNet & SymbolNet & OutNet\\
\hline 
\hline 
Input size & $n_{\text{I0}} = 3n = 12 $ & $ n_{\text{S0}}= 6n = 24$ & $n_{\text{I2}}+n_{\text{S2}}$\\
\hline 
Output size & $n_{\text{I2}}$ & $n_{\text{S2}}$ & $p$\\
\hline 
FC layer & 2 & 2 & 2\\
\hline 
Hidden Nodes & $n_{\text{I1}},n_{\text{I2}}$& $n_{\text{S1}},n_{\text{S2}}$ & $n_{Con},p$\\
\hline 
Output activation & Tanh & Tanh & Sigmoid\\
\hline 
\end{tabular}
\end{table}

Therefore, the final output can be performed as follows
\begin{equation}
    \mathbf{\hat{b}} = f_{\text{Sig}}(\mathbf{W}_{\text{out}}[f_{\text{Tanh}}(\mathbf{W}_1[\mathbf{z}_{\text{index}},\mathbf{z}_{\text{sym}}]+\mathbf{b}_1)]+\mathbf{b}_{\text{out}}),
\end{equation}
where $\mathbf{W}_1$, $\mathbf{W}_{\text{out}}$, $\mathbf{b}_1$, $\mathbf{b}_{\text{out}}$ denote the weights and biases of the two output FC layers.

\vspace{-0.2cm}

\subsection{Model Training\label{subsec:Model Training}}

\vspace{-0.1cm}
Training the DuaIM-3DNet network is implemented by Tensorflow which requires a large amount of training data. Before using the proposed DuaIM-3DNet detector, it is necessary to offline train the DuaIM-3DNet model using the simulation data. To get a corresponding set of transmitted matrix $\mathbf{X} = f_{\text{DM-IM-3D-OFDM}}(\mathbf{b})$, multiple sequences of $p$ bits denoted by $\mathbf{b}$ are randomly generated. Then, under the influence of the Rayleigh fading channel and AWGN, these matrices are transmitted to the receiver. In accordance with their established statistical models, the channel matrix and noise matrix are also randomly created and changed from one bit sequence to another. A block of $\mathbf{Y}$ received signal values is referred to as a data sample. The SymbolNet and IndexNet are learned simultaneously from any data sample. The inputs of the IndexNet and SymbolNet are $(n_{\text{sample}},3n)$ and $(n_{\text{sample}},6n)$, respectively, where $n_{\text{sample}}$ is the sample size.

The DuaIM-3DNet model is trained to reduce the BER, through reducing the discrepancy between sequence $\mathbf{b}$ and its predicted sequence $\mathbf{\hat{b}}$. We use a loss function that is often employed in regression models, namely, mean-squared error (MSE), as follows:
\begin{equation}
    \chi\left(\mathbf{b},\hat{\mathbf{b}};\mathbf{\alpha}\right) = \frac{1}{p} \left \Vert \mathbf{b} - \hat{\mathbf{b}} \right \Vert^2,
\end{equation}
where $\alpha = \left \{[\mathbf{W}_1,\mathbf{W}_{\text{out}}],[\mathbf{b}_1,\mathbf{b}_{\text{out}}] \right \}$ are the weights and biases of model, respectively. We use Adam optimizer \cite{kingma2014adam} and the learning rate is represented by $\eta$.
In order to train DuaIM-3DNet efficiently, the training SNR level (referred to as $\beta_{\text{train}}$) used for training must be chosen properly, since the performance of the trained model is very sensitive to the training SNR. Note that it is crucial to choose the best $\beta_{\text{train}}$ so that the model trained by this training SNR performs well for any other testing SNR of interest.
For instance, the noise impact will not be included in the training if $\beta_{\text{train}}$ is too large, which will result in a poor performance of the DNN model in the testing phase. We will demonstrate the justification for choosing a suitable training SNR based on experiments in Section IV - Simulation Results.

\vspace{-0.2cm}

\subsection{Online Deployment\label{subsec:Online Deployment}}

\vspace{-0.1cm}
After offline training, the DuaIM-3DNet model with optimized parameter $\alpha$ is used for online delivery of detection of DM-IM-3D-OFDM signals with SNRs and channels of interest. The proposed method, in particular, may be used in real-time to estimate data bits over a variety of channel fading situations without the need for extra training for $\alpha$. By feeding the received DuaIM-3DNet signal and the corresponding channel information, our detector quickly and autonomously outputs the predicted bits. 

\section{Simulation results\label{sec:SIMULATION RESULTS}}

\vspace{-0.1cm}

In this section, we provide simulation results in terms of BER and detection runtime. The ML detector is used as the baseline. 
For training, we use a training data size of $5\times10^4$ for each epoch with a batch size of 100 and the number of epochs is set to 500. During training, data samples are randomly generated with each batch. Therefore, we have a total of $25\times10^6$ different data samples.
The learning rate $\eta$ is set to 0.001. Meanwhile, other parameters of DuaIM-3DNet and parameters of DM-IM-3D-OFDM are detailed in Table~\ref{tab:para}. Although the proposed model considers only Rayleigh fading channels in the experiments, the DuaIM-3DNet detector still performs well under other channel models.

\vspace{-0.1cm}

\subsection{BER Performance}

\label{subsec:berPer} 
\begin{figure}[!ht]
    \centering
    \includegraphics[width=\figWidth]{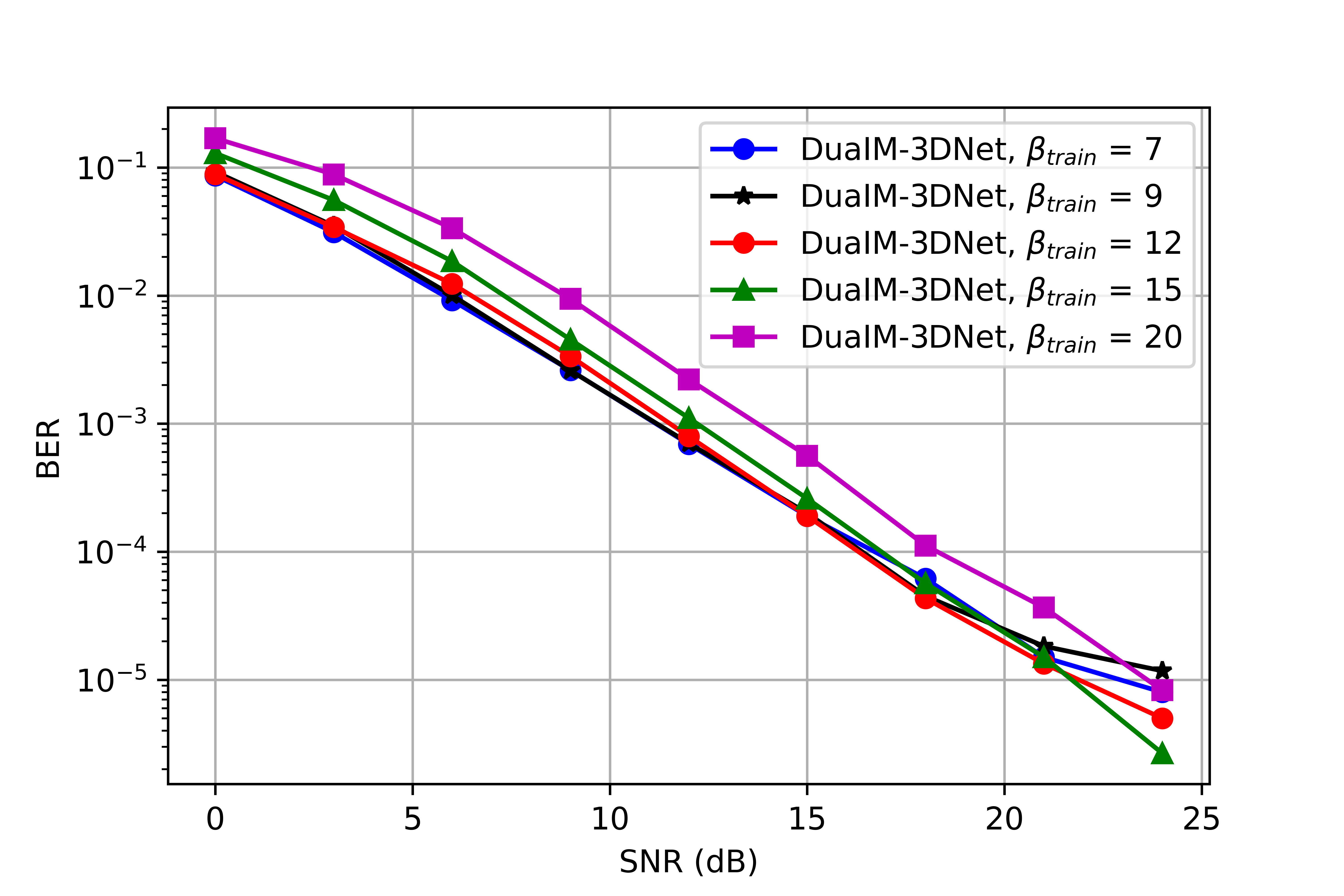}
    \caption[]{BER performance of the proposed DuaIM-3DNet detector when trained with different training SNRs. The system and DNN model parameters for this figure are illustrated in Table~\ref{tab:para}.}
    \label{fig4}
\end{figure}

In Fig.~\ref{fig4}, we demonstrate the BER performance obtained by the DuaIM-3DNet scheme at different training SNRs, where simulation parameters are provided in Table~\ref{tab:para}. It can be seen that the BER performance of our proposed detector significantly varies when changing the training SNR. More specifically, when training with low training SNRs such as 7 dB, our DuaIM-3DNet detector achieves the best BER performance at low SNR regions, while using higher training SNRs such as 15 dB, our proposed DuaIM-3DNet tends to achieve better performance at high SNRs. Moreover, if the training SNR is too large, such as $\beta_{\text{train}}=20\,  \text{dB}$, the BER performance of the proposed DuaIM-3DNet significantly degrades. Based on this observation, in order to achieve the best performance in a wide range of testing SNRs, we propose to train our DNN detector using two training SNRs, namely 7 dB and 15 dB, as shown in the following.

\vspace{-0.1cm}

\begin{figure}[!ht]
    \centering
    \includegraphics[width=\figWidth]{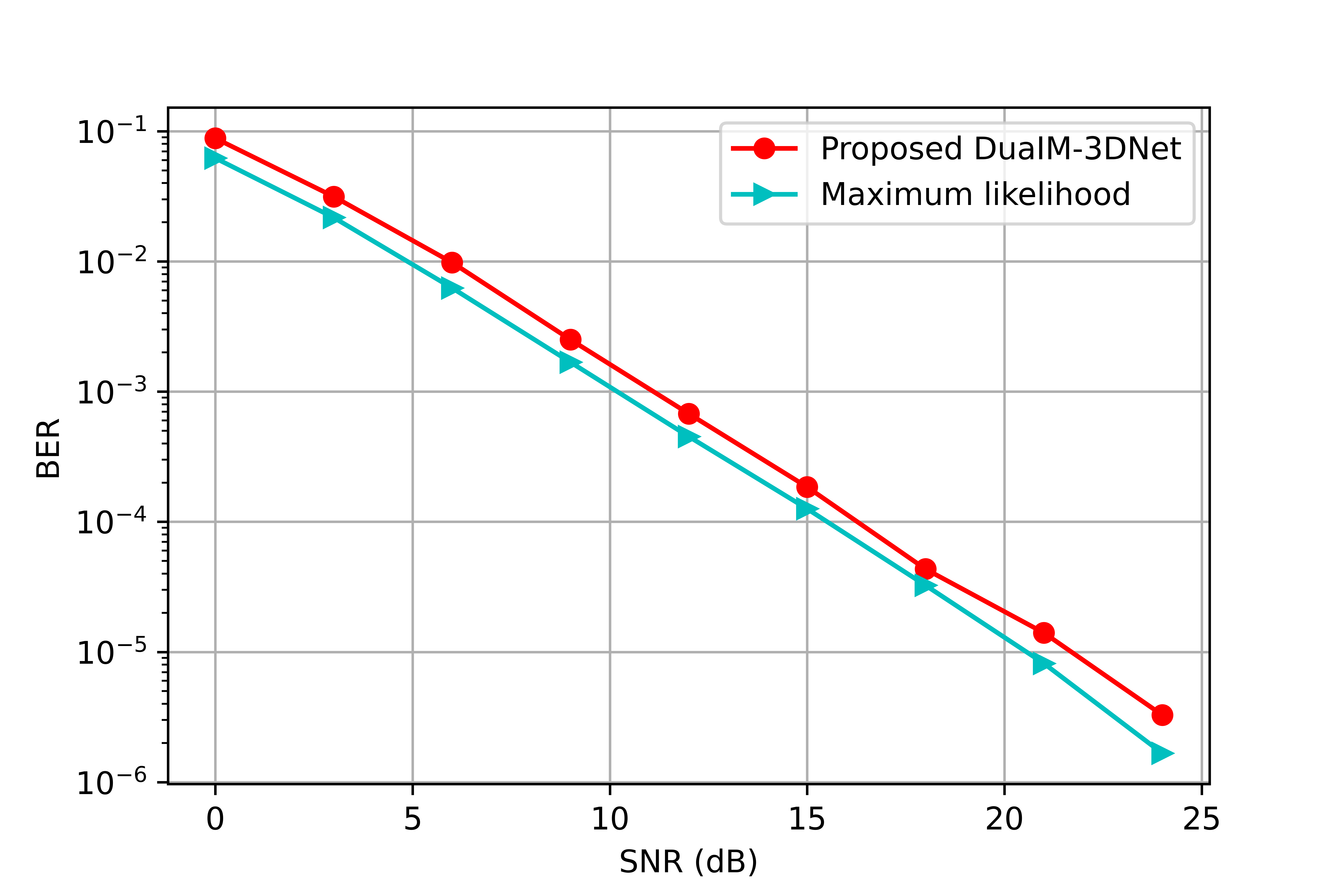}
    \caption[]{BER performance comparison between our proposed DuaIM-3DNet and the ML detector, where our detector is trained with two training SNRs, namely,  $\beta_{\text{train}}^{(1)} = 7$ dB and $\beta_{\text{train}}^{(2)} = $15 dB. The other system and DNN model parameters for this figure are illustrated in Table~\ref{tab:para}.}
    \label{fig5}
\end{figure}

Based on the results observed from Fig.~\ref{fig4}, we found that in order to effectively train the DNN model, we can train our detector with two training SNRs at the same time, denoted as $\beta_{\text{train}}^{(1)}$ and $\beta_{\text{train}}^{(2)}$. More specifically, for each epoch, which is determined by $\text{epoch}\equiv 0 \mod 2$, is trained with $\beta_{\text{train}}^{(1)}$, and the remaining epochs are trained with $\beta_{\text{train}}^{(2)}$. Each training SNR will be trained with the same number of samples, which are different from each other.
We train the model with two training SNRs, namely,  $\beta_{\text{train}}^{(1)} = 7$ dB and $\beta_{\text{train}}^{(2)} = $15 dB. Using such the proposed training strategy, we compare the BER performance between the proposed DuaIM-3DNet and the ML detector in Fig.~\ref{fig5}, where the system parameters of DM-IM-3D-OFDM and the DNN parameters of DuaIM-3DNet are highlighted in Table~\ref{tab:para}. It is shown from Fig.~\ref{fig5} that the proposed DuaIM-3DNet achieves the BER performance very close to that of the ML. More specifically, the performance gap between our detector and the baseline is less than 1 dB across all SNRs.

\begin{table}
\caption{A summary of simulation parameters\label{tab:para}}

\centering{}%
\begin{tabular}{|l|r|}
\hline 
Parameter  & Value\tabularnewline
\hline 
\hline 
Parameters of DM-IM-3D-OFDM $n$, $k$, $c_A$, $c_B$  & 4,2,2,2\tabularnewline
\hline 
Number nodes of layers in IndexNet $n_{\text{I0}}, n_{\text{I1}},n_{\text{I2}}$ & $12,512,256$\tabularnewline
\hline 
Number nodes of layers in SymbolNet $n_{\text{S0}}, n_{\text{S1}},n_{\text{S2}}$  & $24,512,256$\tabularnewline
\hline 
Input size of OutNet $n_{\text{Con}}$  & 512\tabularnewline
\hline 
Number nodes of Tanh activation of OutNet $n_{\text{Tanh}}$  & 256\tabularnewline
\hline 
Output size of OutNet $n_{\text{out}} = p$  & 6\tabularnewline
\hline 
Activation of hidden layers of IndexNet, SymbolNet & \textcolor{black}{Tanh}\tabularnewline
\hline 
Activation of two hidden layers of OutNet & \textcolor{black}{Tanh-Sigmoid}\tabularnewline
\hline 
Training SNR $\beta_{\text{train}}$  & 7,9,12,15,20 dB\tabularnewline
\hline 
Learning rate $\eta$  & 0.001\tabularnewline
\hline 
Batch size  & 100\tabularnewline
\hline 
Number of training epochs  & 500\tabularnewline
\hline 
Training data size  & $25\times10^6$\tabularnewline
\hline 
Testing data size  & $10^{6}$\tabularnewline
\hline 
Optimizer  & Adam \cite{kingma2014adam} \tabularnewline
\hline 
\end{tabular}
\end{table}

\vspace{-0.1cm}

\subsection{Runtime Complexity}

\label{subsec:ComComparision} We transform the DuaIM-3DNet model developed in the Tensorflow environment to the MATLAB environment and compute its runtime to assess the complexity of DuaIM3D-Net. Since the ML detector is also run by the MATLAB using the same computer in order to guarantee a fair comparison. The runtime is measured in seconds for both our DuaIM-3DNet and the ML benchmark. The system parameters and DNN model parameters are illustrated in Table~\ref{tab:para}. The runtime complexity of both detectors is compared in Table~\ref{tab:comlex}. It is shown via this table that the runtime of the ML is $0.001$ seconds, which is significantly higher than that of the proposed DuaIM-3DNet, which consumes only $0.0003$ seconds. This confirms the benefit of our proposed DNN detector in terms of runtime complexity compared the traditional ML detector.

\begin{table}[!ht]
\centering
\caption{Runtime of proposed DuaIM-3DNet and ML detector in seconds\label{tab:comlex}}
\begin{tabular}{|c|c|c|}
\hline
$(n,k,c_A,c_B)$ & ML & DuaIM-3DNet \\ \hline
\hline
(4,2,2,2)           & 0.001    & 0.0003                                             \\ \hline
\end{tabular}
\end{table}

\vspace{-0.2cm}
 
\section{Conclusions\label{sec:Conclusions}}

\vspace{-0.1cm}
In this paper, we proposed the DuaIM-3DNet detector for the DM-IM-3D-OFDM system. In particular, we designed a novel DNN structure for DuaIM-3DNet. Our proposed consists of three sub-neural networks, namely IndexNet for estimating active indices, SymbolNet for estimating 3D-constellation symbols, both relying on the pre-processed data of the received signal and CSI, and then the final DNN called OutNet that combines both IndexNet and SymbolNet for recovering data bits. Once trained with simulated dataset, the proposed detector can be deployed for detecting data bits in an online manner with very low runtime. Our simulation results showed that the proposed DuaIM-3DNet detector can achieve near-optimal BER performance at remarkably lower runtime complexity compared to the conventional ML detector.
 \bibliographystyle{IEEEtran}
\bibliography{IEEEabrv,refs}

\end{document}